\documentclass[aps,prb,showpacs,floatfix,twocolumn,superscriptaddress]{revtex4-1}

\usepackage{graphicx,amsmath,amssymb,bm}
\usepackage{float}
\usepackage{multirow}
\usepackage{hyperref,verbatim}

\begin{document}

\title{Proximity magnetoresistance in graphene induced by magnetic insulators}

\author{D. A. Solis}  
\affiliation{Univ. Grenoble Alpes, CEA, CNRS, Grenoble INP*, IRIG-SPINTEC, 38000 Grenoble, France}
\author{A. Hallal} 
\affiliation{Univ. Grenoble Alpes, CEA, CNRS, Grenoble INP*, IRIG-SPINTEC, 38000 Grenoble, France}
\author{X. Waintal}
\affiliation{Univ. Grenoble Alpes, CEA, IRIG-PHELIQS, 38000 Grenoble, France}
\author{M. Chshiev} 
\affiliation{Univ. Grenoble Alpes, CEA, CNRS, Grenoble INP*, IRIG-SPINTEC, 38000 Grenoble, France}

\begin{abstract}

	We demonstrate the existence of Giant proximity magnetoresistance (PMR) effect in a graphene spin valve where spin polarization is induced by a nearby magnetic insulator. PMR calculations were performed for yttrium iron garnet (YIG), cobalt ferrite (CFO), and two europium chalcogenides EuO and EuS. We find a significant PMR (up to 100\%) values defined as a relative change of graphene conductance with respect to parallel and antiparallel alignment of two proximity induced magnetic regions within graphene. Namely, for high Curie temperature (Tc) CFO and YIG insulators which are particularly important for applications, we obtain 22\% and 77\% at room temperature, respectively. For low Tc chalcogenides, EuO and EuS, the PMR is 100\% in both cases. Furthermore, the PMR is robust with respect to system dimensions and edge type termination. Our findings show that it is possible to induce spin polarized currents in graphene with no direct injection through magnetic materials. 
\end{abstract}


\maketitle
\section{Introduction}

Graphene is a two-dimensional (2D) material \cite{nov2005,kim2005} that has attracted
a lot of interest in view of its unique physical properties
and applications potential in diverse fields such as electronics, spintronics and 
quantum computing \cite{casrev,stephanspintronics,quantumc}. Due to its weak
spin orbit coupling \cite{Tombros-vanWees2007,macdonald2011-prb,Popinciuc-vanWees2009,dlubak-fert2010,han-kawakami2011,yan-ozyilmaz2011,maassen-vanWees2012,dlubak-fert2012,cummings-roche2016,Tuan-roche2016}  graphene possesses a long spin relaxation time and lengths even at room temperature \cite{drogeler}. 
While these characteristics offer an optimal platform for spin manipulation, it remains however a challenge to achieve robust spin polarization efficiently at room temperature.

Several methods have been proposed in 
order to introduce ferromagnetic order on graphene, among which functionalization with adatoms \cite{func2013}, addition of defects \cite{yazyev2007-defts,Yang2011}, and by means of proximity effect via an adjacent ferromagnet\cite{yang2013,Zollnerprb2016,barlas2015-prox,moodera2016,bart2016,brataas2008-prox}. 
The latter approach  attracted a lot of interest using magnetic
insulators (MI) as a substrate to induce exchange splitting in graphene. 
When a material is placed on top of a magnetic insulator, it can acquire 
proximity induced spin polarization and exchange splitting \cite{yang2013} resulting from the
hybridization between $p_z$ orbitals with those of the neighboring magnetic insulator.
For practical purposes, the implementation in spintronic devices of this kind of materials could lead to lower power consumption since no current injection across adjacent ferromagnet (FM) is required as in case of traditional spin injection techniques.
Experimentally, the existence of proximity exchange splitting via magnetic insulator in graphene have been demonstrated with exchange fields up to 100~T using the coupling between graphene and EuS~\cite{moodera2016}.
For yittrium irog garnet/graphene (YIG/Gr) based system, using non-local spin transport measurements, Leutenantsmeyer et al.~\cite{bart2016} demonstrated exchange field strength of ~0.2~T. Another possibility of inducing exchange splitting in graphene using FM metal, by separating them by alternative 2D material such as hexa-boron nitride (hBN), was also proposed theoretically~\cite{Zollnerprb2016}.

Recent studies have suggested the creation of graphene-based devices where EuO-graphene junction can act as a spin filter and spin valve simultaneously by gating the system~\cite{song-spinf-spinv2015}. It was also demonstrated~\cite{song-diode2018} that a double EuO barrier on top of a graphene strip can exhibit negative differential resistance making this system a spin selective diode. However, the drawback of using EuO is its low Curie temperature and the predicted strong electron doping~\cite{yang2013}. It was proposed therefore using high Curie temperature materials such as YIG or cobalt ferrite (CFO)~\cite{ali2017}. Indeed, a large change in the resistance of a graphene-based spintronic device has been reported recently where the heavy doping induced by YIG could be treated by gating~\cite{Song-gating}. 
 
In this Letter we demonstrate the existence of Proximity Magnetoresistance (PMR) effect in graphene for four different magnetic insulators (MI), YIG, CFO, europium oxide (EuO) and europium sulfide (EuS). Using ab initio parameters reported in Ref.~[\citenum{ali2017}], we show that for YIG and CFO based lateral graphene-based devices with armchair edges, PMR values could reach 77\% and 22\% at room temperature (RT), respectively. With chalcogenides, EuS and EuO, PMR values can reach 100\% at 16 K and 70 K, respectively. In addition, we demonstrate the robustness of this effect with respect to system dimensions and edge type termination. Furthermore, our calculations with spin-orbit coupling (SOC) included does not significantly affect the PMR. These findings will stimulate experimental investigations of the proposed phenomenon PMR and development of other proximity effect based spintronic devices.
\section{Methodology}

\begin{figure}[t]
	\centering
	\includegraphics[width=0.8\columnwidth]{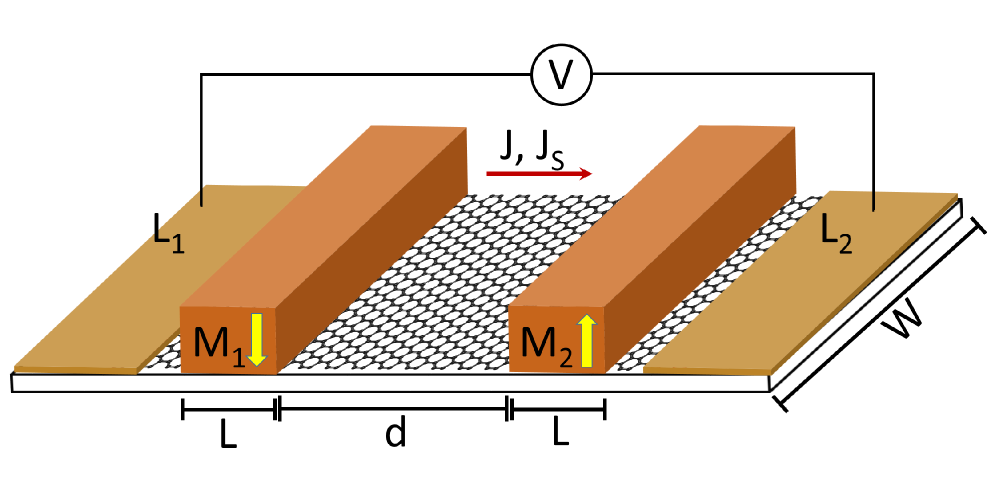}
	\caption{(color online) Lateral spintronic device comprising two magnetic insulators on top of a graphene sheet. The magnetic graphene regions have a length $L$, width $W$ and are separated by a distance $d$.}
	\label{fig_1}
\end{figure}

In order to calculate conductances and PMR, we employed the tight-binding approach with scattering matrix formalism conveniently implemented within the KWANT package~\cite{xavier}. The system modeled is shown in Fig.~\ref{fig_1} and comprises two identical proximity induced magnetic regions of width $W$ and length $L$ resulting from insulators with magnetizations {\bf M}$_1$ and {\bf M}$_2$, separated by a distance $d$ of nonmagnetic region of graphene sheet with armchair edges. Both magnetic graphene regions are separated from the leads  $L_1$ and $L_2$ by a small pure graphene region. In order to take into account the magnetism arising in graphene from the proximity effects induced by the MI's, in the Hamiltonian are used the parameters obtained for different MI's in Ref. ~[\citenum{ali2017}]. It is important to note that the magnetic regions do not affect the linear dispersion of graphene bands, except breaking the valley and electron-hole symmetry resulting in spin-dependent band splitting and doping. The discretized Hamiltonian for the magnetic graphene regions can be expressed as:
\begin{widetext}
\begin{equation}\label{eq:1}
H = \sum_{i\sigma} \sum_l t_{l\sigma} c^\dagger_{(i+l)1\sigma} c_{i0\sigma} +h.c.
+  \sum_{i\sigma\sigma'}\sum_{\mu=0}^1 
\left[\delta+(-1)^\mu\Delta_\delta\right] c^\dagger_{i\mu\sigma} [\vec m . \vec\sigma] c_{i\mu\sigma'} 
+  \sum_{i\sigma}\sum_{\mu=0}^1 
\left[E_D+(-1)^\mu\Delta_s\right] c^\dagger_{i\mu\sigma} c_{i\mu\sigma} 
\end{equation}
\end{widetext} 
where $c^\dagger_{i\mu\sigma}$ ($c^\dagger_{i\mu\sigma}$) creates (annihilates) an electron of type $\mu=0$ for A sites and $\mu=1$ for B sites on the unit cell $i$ with spin $\sigma=\uparrow(\downarrow)$ for up (down) electrons.
$\vec m$ and $\vec\sigma$ respectively represent a unit vector that points in the direction of the magnetization and the vector of Pauli matrices, so that $\vec m.\vec\sigma = m_x \sigma_x+m_y\sigma_y+m_z\sigma_z$.
The anisotropic hopping $t_{l\sigma}$ connects unit cells $i$ to their nearest neighbor cells $i+l$. 
Parameters $\delta$, $\Delta_\delta$, $\Delta_s$ are defined via exchange spin-splittings ${{\delta}_{e}}$ (${{\delta}_{h}}$) of the electrons (holes) and spin-dependent band gaps $\Delta_{\sigma}$ defined in Ref.~[\citenum{ali2017}]. $E_D$ indicates the Dirac cone position with respect to the Fermi level. The Hamiltonian for the whole device is obtained by making aforementioned parameters spatially dependent.

\begin{figure*}[ht]
	\centering
	\includegraphics[width=1\linewidth]{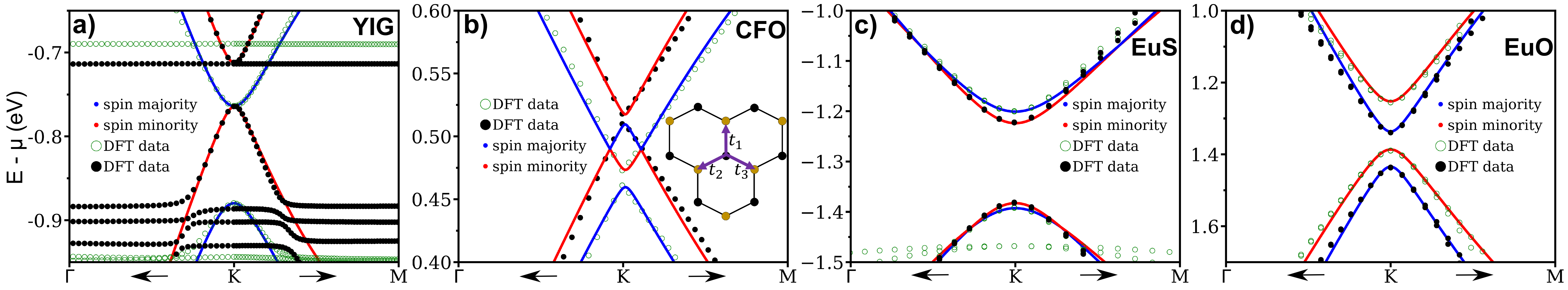}
	\caption{(color online) Band structure obtained using tight-binding Hamiltonian defined by Eq.~(\ref{eq:1}) (solid lines) fitted to the band structure from DFT spin majority (green open circles) and spin minority (black filled circles) data for the cases with (a) YIG, (b) CFO, (c) EuS and (d) EuO from Ref.~[\citenum{ali2017}]. The inset in (b) shows the anisotropic hoppings reported in Table~\ref{tab1}}.
	\label{fig_2}
\end{figure*}

To obtain hopping parameters of Hamiltonian~(\ref{eq:1}), we fitted tight-binding bands to those obtained from first principles calculations in Ref.~[\citenum{ali2017}]. The results of the fitting procedure in case of graphene magnetized by YIG, CFO, EuS and EuO are shown in Fig.~\ref{fig_2}(a), (b), (c) and (d), respectively. The corresponding hopping parameters are given in Table~\ref{tab1}. As one can see, the graphene bands obtained with tight-binding Hamiltonian given by Eq.~\ref{eq:1} are in good agreement with those obtained using Density Functional Theory (DFT) confirming suitability of our model for transport calculations. Of note, due to the presence of superficial tension at the interface between CFO and graphene, hopping parameters in this case are anisotropic as they depend on direction to the nearest neighbor as specified in the inset of Fig.~\ref{fig_2}(b).

\begin{table}[ht]
\caption {hopping parameters used in equation \ref{eq:1} for each magnetic insulator considered.} 
\label{tab1} 
{\footnotesize
\begin{center}
\begin{tabular}{ p{1.6cm}|p{1.7cm}|p{2cm}|p{2.2cm} }
\hline
\hline
Material &  Hopping direction & spin up (eV) & spin down (eV)\\
\hline
\hline
YIG & t & 3.6   & 3.8 \\
\hline
\multirow{3}{4em}{CFO} & $t_1$ & 1.38             & 1.44            \\ 
                        & $t_2$ & $1.41e^{-i0.01}$ & $1.48e^{-i0.01}$ \\ 
                        & $t_3$ & $1.36e^{-i0.02}$ & $1.44e^{-i0.02}$ \\ 
\hline                
EuS & t & 4.5 & 4.8 \\
\hline
EuO & t & 4.9 & 4.3 \\
\hline
\end{tabular}
\end{center}
}
\end{table}

The conductance for parallel and antiparallel configurations of magnetizations  {\bf M}$_1$ and {\bf M}$_2$ in the linear response regime is then obtained according to: 

\begin{equation}
G_{P(AP)}=\frac{e}{h}\sum_{\sigma}\int T_{P(AP)}^\sigma \left(\frac{-\partial f}{\partial E} \right)dE ,
\label{eq:1b}
\end{equation}
where $T_{P(AP)}^\sigma$ indicates spin-dependent transmission probability for parallel(antiparallel) magnetizations configurations and $f=1/(e^{(E-\mu)/k_BT}+ 1)$ represents the Fermi-Dirac distribution with $\mu$ and $T$ being electrochemical potential (Fermi level) and temperature, respectively. It is important to note that temperature smearing has been taken into account using the Curie temperature of each MI.

The PMR amplitude has been defined according to following expression:
\begin{equation}
\textrm{PMR}=\left(\frac{G_P -G_{AP}}{G_P +G_{AP}}\right)\times 100 \% ,
\label{eq:2}
\end{equation}
In order to determine the impact of the system dimensions on the PMR, several calculations were carried out for different lengths, widths and separations of the magnetic regions. Furthermore, we checked the robustness of PMR on edge type termination by calculating the PMR for systems with zigzag,  armchair and rough edges. The latter were created by removing atoms and bounds randomly and deleting the dangling atoms at the new edges.

\section{Results}

\begin{figure}[b]
	\centering
	\includegraphics[width=1\columnwidth]{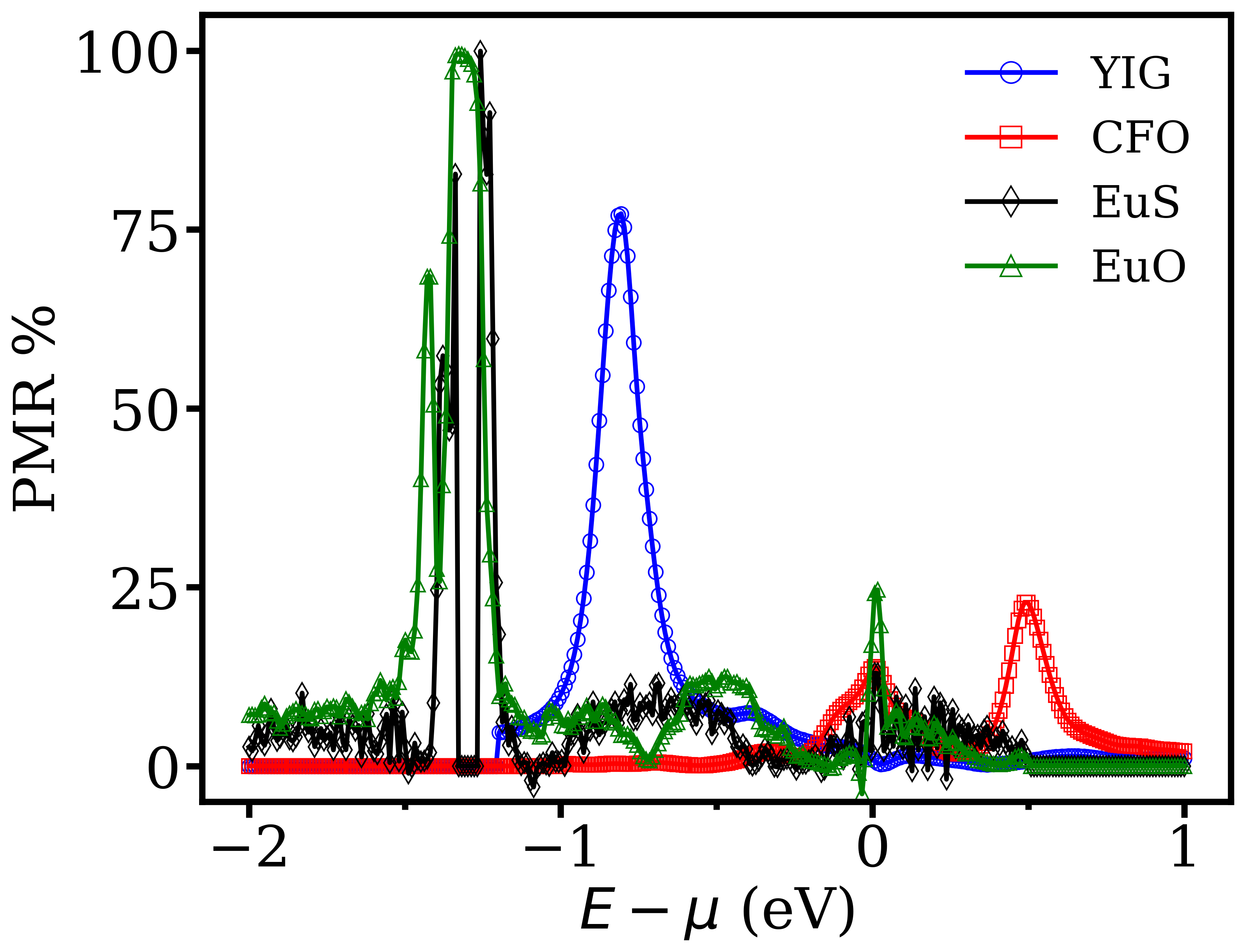}
	\caption{(color online) Proximity magnetoresistance defined by Eq.~\ref{eq:2} as a function of energy in respect to the Fermi level for YIG (blue circles), CFO(red squares), EuS(black diamonds) and EuO(green triangles) using temperature smeared conductances at T=300 K, 300 K, 16 K and 70 K, respectively. System dimensions are $L=49.2$~nm, $W=39.6$~nm and $d=1.5$~nm. }
	\label{fig:fig_3}
\end{figure}

In Fig.~\ref{fig:fig_3} we present the PMR curves for lateral device structures based on YIG, CFO, EuS and EuO on top of a graphene sheet with armchair edges. Taking into account Curie temperatures for these materials, the curves were smeared out using 16 K (70 K) for EuS (EuO), and 300 K for YIG and CFO cases. 
For system with YIG we found a maximum PMR value of 77\% while for CFO the value obtained was 22\%. In case of chacolgenides EuS and EuO used, the maximum PMR values reach 100\%. Among the materials studied, YIG represents the most suitable candidate for lateral spintronic applications due to both high Curie temperature and considerably large PMR value.

\begin{figure*}[t]
	\centering
	\includegraphics[width=1\linewidth]{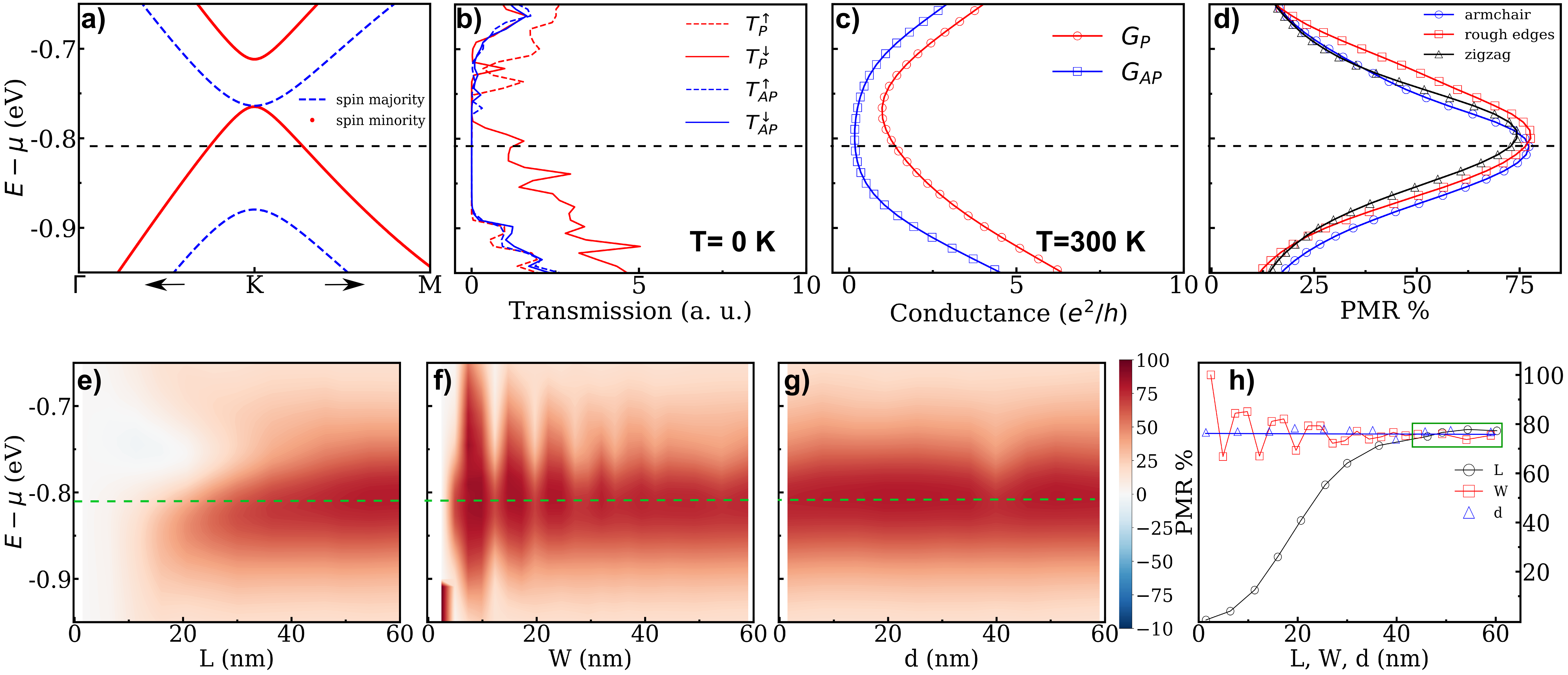}
	\caption{(color online) (a) Band structure reproduced using the DFT parameters from Ref.~[\citenum{ali2017}] for graphene in proximity of YIG. (b) Transmission probabilities for majority (dash lines) and minority (solid) spin channel for parallel (red) and antiparallel (blue) magnetization configurations at $T=0$ K for a system with dimensions $L=49.2$~nm, $W=39.6$~nm and $d=1.5$~nm. (c) Resulting conductance for parallel (red circles) and antiparallel (blue squares) magnetization configurations at 300~K. (d) PMR for device with armchair (blue circles), rough (red squares) and zigzag (black triangles) edge termination of graphene. PMR profiles as a function of (e) $L$, (f) $W$ and (g) $d$. (h) Dependence of PMR for the energy outlined by dashed line in (e), (f) and (g) as a function of $L$ (black circles), $W$ (red squares) and $d$ (blue triangles). The green square highlights the region where PMR becomes independent of system dimensions.}
	\label{fig:fig_4}
\end{figure*}

In order to elucidate the underlying physics behind these PMR results, let us analyze details of the conductance behaviour. In Fig.~\ref{fig:fig_4}(a)-(b) we reproduce the graphene bands in proximity of YIG and corresponding transmission probabilities resolved in spin for P and AP configurations at $T=0$ K for a system with dimensions $L=49.2$ nm, $W=39.6$ nm and $d=1.5$  nm.  One can see that for energies between -0.88 eV and -0.78 eV there is no majority spin states present and the only contribution to transmission $T_{P}^\downarrow$ is from minority spin channel (Fig.~\ref{fig:fig_4}(b), red solid line). In other words, the situation within this energy range is half-metallic giving rise to maximum PMR values of 100\% using ``pessimistic" definition given by Eq.~(\ref{eq:2}). The similar situation is for energy ranges between -0.72 eV and -0.75 eV but this time the only contribution $T_{P}^\uparrow$ is from majority spin channel (Fig.~\ref{fig:fig_4}(b), red dashed line). One should point out here that the conduction profile here is due combining both magnetic and nonmagnetic regions into one scattering region. The conductance of a pure graphene nanoribbon sheet represents quantized steps due to transverse confinement with no conductivity at zero energy depending on its edges. Inducing magnetism within graphene sheet leads to symmetry breaking with the shift of exchange splitted gaps in the vicinity of Dirac cone region below the Fermi level. This leads to characteristic conductance profile with two minima at around -0.8~eV and 0~eV (not shown here) due to the Dirac cone regions of magnetized and the pure graphene. 
The corresponding conductances for the parallel ($G_P$) and for the antiparallel ($G_{AP}$) magnetic configurations at $T=300$ K are shown in Fig~\ref{fig:fig_4}(c). Interestingly, even at room temperature the PMR for YIG based structure preserves a very high value of about 77\% as already pointed above, a behavior that is very encouraging for future experiments on PMR. As a guide to the eye with dashed lines we highlight the energy value where the PMR has a maximum in Fig. \ref{fig:fig_4}.

Since the edges may strongly influence the aforementioned properties of the system, we next explore the robustness of PMR against different edge types of the graphene channel of the proposed device. It is well known that electric field can trigger half-metallicity in zigzag nanoribbons due to the antiferromagnetic interaction of the edges~\cite{zz-loui}. On the other hand, graphene nanoribbons with armchair edges can display insulating or metallic behaviour depending on graphene nanoribbon (GNR) width~\cite{dressel-arm,exp-armc-peter}. Armchair and zigzag edges are particular cases and the most symmetric edge directions in graphene. But one can cut GNR at intermediate angular direction between these two limiting cases giving rise to an intermediate direction characterized by a chirality angle~$\theta$\cite{oleg2013-review}.
Graphene band structure is highly dependent on $\theta$. When the angle is increased, the length of the edge states localized at the Fermi level decrease and eventually disappear in the limiting case when $\theta =30^{\circ}$, i. e. 
when acquires armchair edge. In the laboratory conditions, graphene sheets are finite and have
imperfections that influence their transport properties. For defects at the edges, it has been demonstrated that rough edges can diminish the conductance of a graphene 
nanoribbon as was shown in Ref.~[\citenum{libisch2012-rough}] or may exhibit a nonzero spin conductance as reported in Ref.~[\citenum{wimmer2008-rough}].

In order to demonstrate the robustness of PMR with respect to the edge type, we thus performed calculations with the same system setup (Fig.~\ref{fig_1}) but this time for various edge terminations. The resulting PMR behavior for the cases with armchair, rough edges and zigzag are shown in Fig.~\ref{fig:fig_4}(d). The former have been modeled by creating extended vacancies distributed randomly. It is clear that the maximum PMR value does not present a significant variation maintaining for all cases PMR values around 75\%. With this results in hand we can claim that the PMR is indeed robust with respect to edge termination type.

As a next step, we checked the dependence of the PMR on different system dimensions, i.e. the length of the magnetic region $L$, system width $W$ and the separation between the magnetic regions $d$. The corresponding dependences are presented respectively in Fig.~\ref{fig:fig_4}(e),(f) and (g). One can see that for all energy ranges the PMR ratio has a tendency to increase as a function of $L$ approaching limiting value of 77\% at energies around -0.81 eV indicated by a dashed line Fig.~\ref{fig:fig_4}(e). As for dependence of the PMR as a function of GNR width $W$, clear oscillations  due to quantum well states formation are present with a tendency to vanish as system widens (Fig.~\ref{fig:fig_4}(f)). On a contrary, the PMR shows almost constant behavior as a function of separation between the magnets $d$ (Fig.~\ref{fig:fig_4}(g)) due to the fact that transport is in ballistic regime. For convenience, we summarize all these dependencies in Fig.~\ref{fig:fig_4}(h) at energy -0.81 eV as a function of $L$, $W$ and $d$. One can clearly see that the PMR saturates as system dimensions are increased. At the same time, it shows the oscillations in the PMR for small $W$ as well as the invariance of the PMR with respect to $d$. For large dimensions highlighted by the green box in Fig.~\ref{fig:fig_4}(h), we can claim that the PMR is indeed robust, and the maximum PMR value would be eventually limited only by the magnitude of the spin diffusion length in the system. 

\begin{figure}[t]
\centering
\includegraphics[width=1\linewidth]{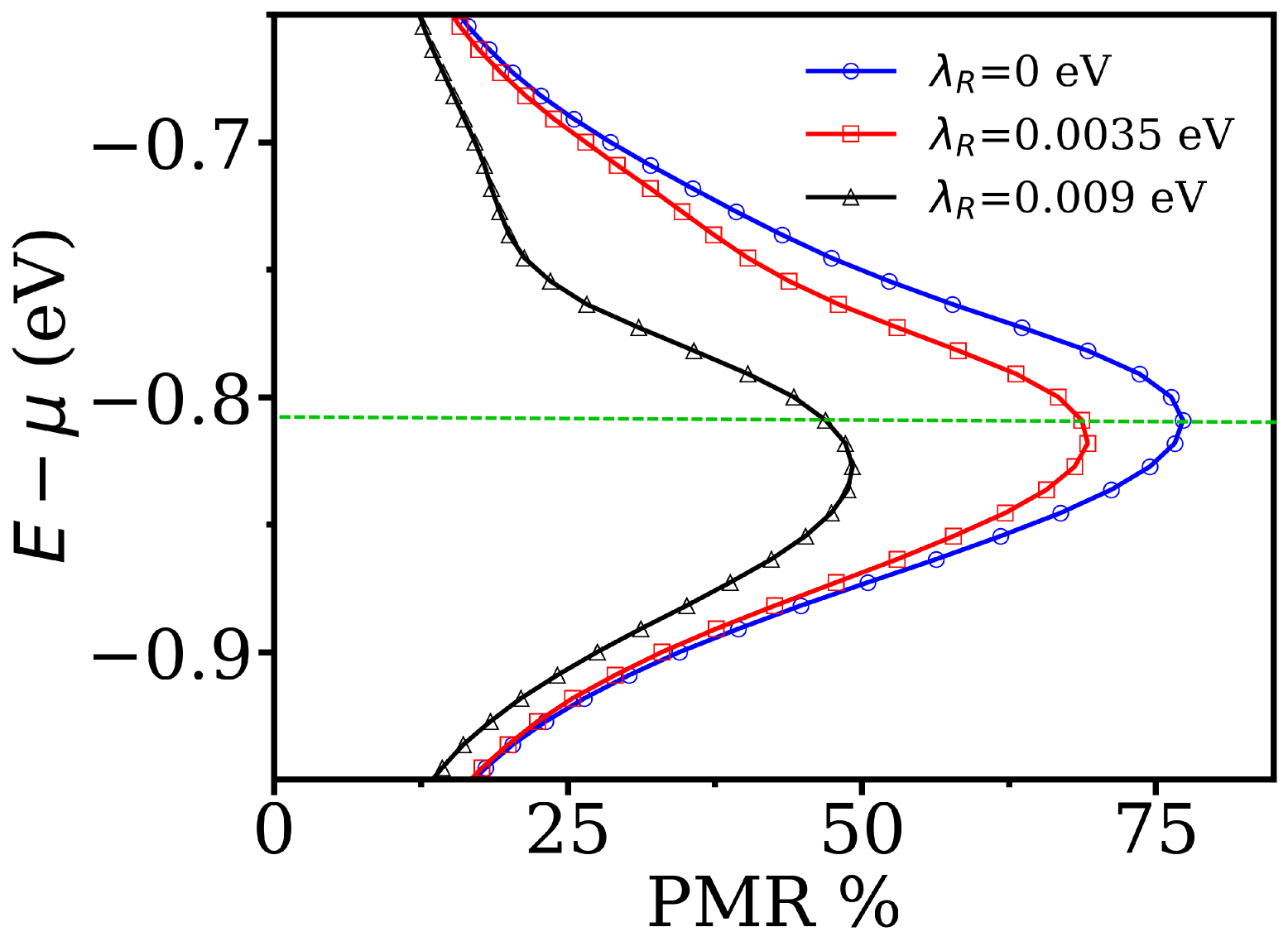}
\caption{(color online) PMR dependencies for three values of Rashba spin-orbit interaction parameter $\lambda_R$ defined by Eq.~(\ref{soc}) for YIG-based system with armchair edges and of dimensions $L=49.2$~nm, $W=39.6$~nm and $d=1.5$~nm. The dashed line is a guide to the eye that shows the maximum value when $\lambda_R=0$~eV.}
\label{fig:fig_5}
\end{figure}

Finally, we consider the impact of spin-orbit coupling on the PMR. Despite weak SOC within graphene, the proximity of adjacent materials can induce the interfacial Rashba SOC~\cite{macdonald2011-prb}. Rashba type SOC is included into our tight-binding approach adding the following term:

\begin{equation}\label{soc}
H_{SO} = i \lambda_R \sum_{i\sigma\sigma'} \sum_l  
c^\dagger_{(i+l)1\sigma} [\sigma^x_{\sigma\sigma'} d_l^x - \sigma^y_{\sigma\sigma'} d_l^y] c_{i0\sigma'} +h.c.
\end{equation}
where the vector $\vec d_l=(d_l^x,d_l^y)$ connects the two nearest neighbours, $\lambda_R$ indicates the SOC strength. The values of $\lambda_R$ are generally lie in the range between ~1-10~meV (see, for instance, in Ref.~[\citenum{macdonald2014-prl}]). Keeping in mind this information, we present in in Fig.~\ref{fig:fig_5} the PMR dependences for three values of spin-orbit interaction. One can see that increasing the strength of SOC $\lambda_R$ lower the PMR. This behavior is expected and could be attributed to the fact that spin-orbit interaction mixes the spin channels. These dependencies allows us to conclude that PMR is quite robust also against SOC and even in the worst scenario remains of the order of 50~\% (cf. black triangles and blue circles in Fig.~\ref{fig:fig_5}).

\section{Conclusions}
In this paper we introduced the proximity induced magnetoresistance phenomenon in graphene based lateral system comprising regions with proximity induced magnetism by four different magnetic insulators. For YIG and CFO based devices we found PMR ratios of 77\% and 22\% at room temperature, respectively. For chalcogenide based systems, i.e. with EuS and EuO, we found PMR values of 100\% for both at 16 K and 70 K, respectively. Very importantly, it is demonstrated that the PMR is robust with respect to system dimensions and edge type termination. Furthermore, the PMR survives in case of the presence of SOC decreasing only by about a half even in the case of considerably big SOC strength values. We hope this work will encourage further experimental research and will be useful for the development of novel generation of spintronic devices based on generation and exploring spin currents without passing charge currents across ferromagnets.

\section{acknowledgments}
We thank J. Fabian and S. Roche for fruitful discussions. This project has received funding from the European Union’s Horizon 2020 research and innovation programme under grant agreements No. 696656 and 785219 (Graphene Flagship). X.W. acknowledge support by ANR Gransport.


\appendix



\bibliography{reference}

\end{document}